# Taxonomy, Semantic Data Schema, and Schema Alignment for Open Data in Urban Building Energy Modeling

Liang Zhang[1,2], Jianli Chen[3], Jia Zou[4]
1. University of Arizona, Tucson, Arizona, U.S.
2. National Renewable Energy Laboratory, Golden, Colorado, U.S.
3. University of Utah, Salt Lake City, Utah, U.S.
4. Arizona State University, Phoenix, Arizona, U.S.

* Corresponding author:
Email address: liangzhang1@arizona.edu
Postal address: 1209 E 2nd St, Tucson, AZ 85719, U.S.

## Abstract
Urban Building Energy Modeling (UBEM) is a critical tool to provide quantitative analysis on building decarbonization, sustainability, building-to-grid integration, and renewable energy applications on city, regional, and national scales. Researchers usually use open data as inputs to build and calibrate UBEM. However, open data are from thousands of sources covering various perspectives of weather, building characteristics, etc. Besides, a lack of semantic features of open data further increases the engineering effort to process information to be directly used for UBEM as inputs. In this paper, we first reviewed open data types used for UBEM and developed a taxonomy to categorize open data. Based on that, we further developed a semantic data schema for each open data category to maintain data consistency and improve model automation for UBEM. In a case study, we use three popular open data to show how they can be automatically processed based on the proposed schematic data structure using large language models. The accurate results generated by large language models indicate the machine-readability and human-interpretability of the developed semantic data schema.

## 1. Introduction

The role of buildings in carbon dioxide emissions is more significant than often appreciated, with over a third of all U.S. emissions attributable to this sector - a share exceeding that of the entire industrial sector [1]. The importance of understanding and addressing this contribution has led to the evolution of Urban Building Energy Modeling (UBEM) [2]. UBEM harnesses an array of methodologies and tools to forecast energy demand, identify peak load periods, and analyze energy consumption patterns of building stocks at various scales - from urban to nation, and to global. By facilitating a deeper understanding of energy use in buildings, UBEM provides crucial insights for strategies aimed at reducing carbon emissions and advancing sustainable urban development.

In response to the escalating need for energy modeling at the urban scale, bottom-up, physics-based UBEM [3] has become a central research focus. This bottom-up approach is characterized by the granular computation of energy consumption for each individual building or representative samples. This is then aggregated to provide an estimate at the city, regional, or national level. The modeling methodology employed in this approach is based on physics-based models that utilize a first-principle simulation engine. These models consider a variety of input factors, including weather conditions, structural characteristics of the building, building systems, and equipment, to precisely calculate the energy consumption of buildings. Bottom-up physics-based UBEM tools, such as Comstock [4] and CityBES [5], offer the advantage of generating high-resolution spatiotemporal energy simulation results for building end-use. This level of detail enhances the accuracy and reliability of the models, making them invaluable tools in the quest for sustainable urban development.

## 1.1 Data and Open Data for UBEM

Despite its immense value, UBEM is posed with significant challenges especially in the data perspective [6]. Socio-technological data are critical input of UBEM. However, much of the existing research in the realm of UBEM predominantly focuses on the modeling aspects; there appears to be a dearth of comprehensive studies that delve into the exploration of data for UBEM. In terms of data challenges, ***first, the process of data collection for UBEM is a daunting task that demands substantial time, resources, and expertise.*** The task of modeling the energy usage of thousands to millions of buildings necessitates the collection and processing of a vast array of data. This includes detailed information about the buildings' characteristics, their geometric parameters, local weather data, and other socio-technological variables. The sheer volume of necessary data, coupled with the complexity of the models, places a significant burden on UBEM modelers. They must not only gather this information but also ensure its accuracy and relevance.

***Second, open data is essential to the data perspective of UBEM, but systematic research in open data for UBEM is lacking.*** In the era of information proliferation, open data – freely accessible, usable, modifiable, and shareable data sets – has emerged as a crucial resource for various fields, including UBEM [7]. The significance of open data for UBEM lies in the diverse array of social-technological information it provides. These data sets offer invaluable inputs that can be leveraged to enhance the accuracy and depth of UBEM. Open data sources are not static; they are expanding rapidly, both in terms of volume and variety. This dynamic growth ensures a continuous stream of fresh, relevant data to support UBEM. From detailed building characteristics to localized weather patterns, the open data provides essential resources to building and urban scientists, facilitating more comprehensive and precise modeling, and thereby contributing significantly to the field's advancements. Open data applicable to UBEM can be procured from a myriad of sources and may manifest in a multitude of forms. Such data can be sourced from building footprint datasets, which provide essential details about a building's location and physical structure, and land use datasets, which offer insights into the environmental context of the building. Other crucial sources of open data include assessor recorder datasets, energy disclosure datasets, and building energy certificate databases. Furthermore, data from surveys and building permit records can provide additional valuable context, while local building energy codes and standards offer crucial regulatory information. On-site surveys and local databases, meanwhile, often serve as rich sources of localized and granular data. Even open commercial datasets, which may offer insights into building usage patterns and occupancy behaviors, can be of significant value.

Goy et al. [8] which presents a joint review of data impact and data accessibility to highlight areas for future survey efforts. Besides, Chen et al. [7] introduce data needs, data standards, and data sources required to develop city building datasets for UBEM. The authors conducted a literature review of data needs for UBEM and reviewed the capabilities of current data standards for city building datasets. They also studied existing public data sources from several pioneer cities to evaluate their adequacy in supporting UBEM. The results of the study show that most cities have adequate public data to support UBEM. However, the data are represented in different formats without standardization, and there is a lack of common keys to make the data mapping easier. Malhotra et al. [9] mentioned in their literature review of information modeling of UBEM that future developments should focus on the enrichment of open data sets and on storing the information as common data formats such as gbXML and CityGML Energy ADE. The highlighted key findings are: (1) the majority of papers utilize open data as inputs for UBEM [10], (2) the types of open data used are often limited to a few categories [8], (3) excessive assumptions are made in the absence of available data [11], (4) there is a recurring pattern of finding and processing open data across different papers [12], and (5) UBEM routinely encounters the typical challenges of data availability and granularity [13]. These findings motivated several contributions made in this work.

## 1.2 Research Scope

**(1) A systematic review and taxonomy of open data applicable to UBEM.**

To systematically summarize open data, it's paramount to comprehend diverse types of open data benefiting UBEM and the manner in which these data types are classified. Chen et al. [7] highlighted the necessity of amassing building asset data on a city-wide scale from diverse sources, including surveys, city projects, public records, and city datasets. Existing taxonomies, both formal and informal, have been developed for data within the context of UBEM. For instance, Goy et al. [8] have categorized open data for UBEM into four distinct groups: 1) Census data, 2) 2D and 3D maps, 3) On-site surveys / Local databases, and 4) Building energy certificate databases. Reinhart and Davila [2] have divided data for UBEM into two categories: 1) weather information and 2) building information. Chen et al. [7] have categorized open data for UBEM into three groups: 1) geometry, 2) segmentation parameters, and 3) energy use data. Despite these existing categorizations, a dedicated taxonomy that takes into account the specific context of open data within UBEM is largely unexplored. Malhotra et al. [9] categorize UBEM input into 7 categories: location and geometry, openings, thermal zones and thermal boundaries, building physics, building systems, usage, and internal heat gains. However, the categorization or taxonomy of these papers just focuses on the perspective of general data for UBEM, instead of highlighting open data for UBEM. A systematic taxonomy of open data will not only provide modelers with a broader spectrum of information for their modeling tasks, but also promote a more methodical approach to the use of open data, as opposed to arbitrary data search and selection.

**(2) A novel semantic data schema/structure to facilitate the utilization of open data for UBEM**.

Researchers (e.g., Chen et al. [7]) advocated for the assembly of data for UBEM into a unified open database, employing standardized formats and terminologies. While their work involved standardizing the dataset used for their particular case study, the standardization process they adopted was largely specific to that case and may not be universally applicable. This highlights the prevailing gap in research: the lack of a comprehensive study aimed at standardizing and summarizing open data in a manner that transcends case-specific limitations. Such a study could greatly enhance the usability and accessibility of open data for UBEM, irrespective of the specificities of individual case studies.

Semantic data schemas provide a standardized way to describe the meaning, relationships, and nuances of data elements, ensuring that they can be easily understood, retrieved, and utilized across various platforms and applications. Implementing a well-defined semantic structure is paramount for a couple of key reasons. Firstly, it ensures data interoperability, allowing disparate datasets from different sources to be combined and analyzed coherently. Given that UBEM often requires collating data from multiple sources, a shared semantic structure ensures that data points from one source can be related seamlessly to another, facilitating holistic modeling. Secondly, semantic structures offer a degree of future-proofing. As the domain of UBEM evolves, the data needs will change, but a robust semantic structure can adapt to these changes, ensuring that newer datasets can be integrated without major overhauls. Thirdly, it helps in reducing the ambiguity and variability in data interpretation, which are common pitfalls when diverse datasets are amalgamated without a common schema. Moreover, with the emergence of Large Language Models (LLMs) like GPT variants, there's an amplified potential to fully automate the schema alignment process with the semantic data schema. LLMs, equipped with their profound comprehension of context and semantics, can adeptly identify, map, and integrate diverse data elements into the established schema. As such, the foundational significance of a robust semantic data schema is further magnified, serving as the cornerstone for future LLM-based schema alignment endeavors. This convergence of advanced AI techniques with structured semantic frameworks heralds a new era of efficient and accurate data integration for UBEM and beyond.

A standard semantic data schema can be central to optimize the use of open data for UBEM to address the challenges of data availability and granularity, as well as the need to transcend case-specific limitations. Such an endeavor would not only streamline the data collation and modeling processes but also ensure that insights derived are consistent, reliable, and universally applicable. Although semantic data schema has been discussed in the general field of urban science [14] [15] [16] [17], the semantic data schema especially designed for UBEM is missing in the literature.

### 1.3 Paper Organization

In this paper, based on an extensive review, we first aim to construct a comprehensive taxonomy that categorizes open data, thereby offering a more structured approach to their usage. Building upon this taxonomy, we propose to develop domain-knowledge-based semantic data schema for each category of open data. This not only ensures data consistency but also enhances the level of automation achievable in UBEM. Based on the schema, we automatically align real open data to the schema with LLM in case studies to demonstrate its machine readability and human interpretability. By presenting a more systematic, structured, and automated approach, we aspire to contribute to the efficient utilization of open data in the field of UBEM.

The structure of this paper is organized as follows: Section 2 presents a detailed introduction to our proposed taxonomy for open data in the context of UBEM. Following this, we develop the semantic data schema that we have proposed for each category of open data in Section 3. We also conduct three case studies to automatically align three well-known open data to the developed schema with LLMs to demonstrate machine-readability and human-interpretability of the schema. Finally, Section 5 encapsulates our key findings, and outlines potential future directions for data collection and the continued evolution of open data in this domain.

## 2. Taxonomy of Open Data for UBEM

In this paper, we introduce a taxonomy specifically tailored to the context of open data for UBEM. Originating from the inputs to UBEM, as demonstrated in Figure 1, and akin to the energy modeling for a single building, UBEM consists of seven categories of inputs: (1) weather information derived from weather stations and services, (2) building stock geographic information system (GIS) information depicting the geometries of thousands, potentially millions, of buildings, (3) occupant behavior information detailing building operations, occupancy, and behaviors, (4) building characteristics and energy systems, (5) macroscopic building energy data which is critical for calibrating the UBEM model on a large scale, (6) microscopic building energy data which is vital for calibrating the UBEM model at the single building scale, and (7) a comprehensive data lake that amalgamates multiple data sources as listed above. Each of these groups is discussed in detail within this section.

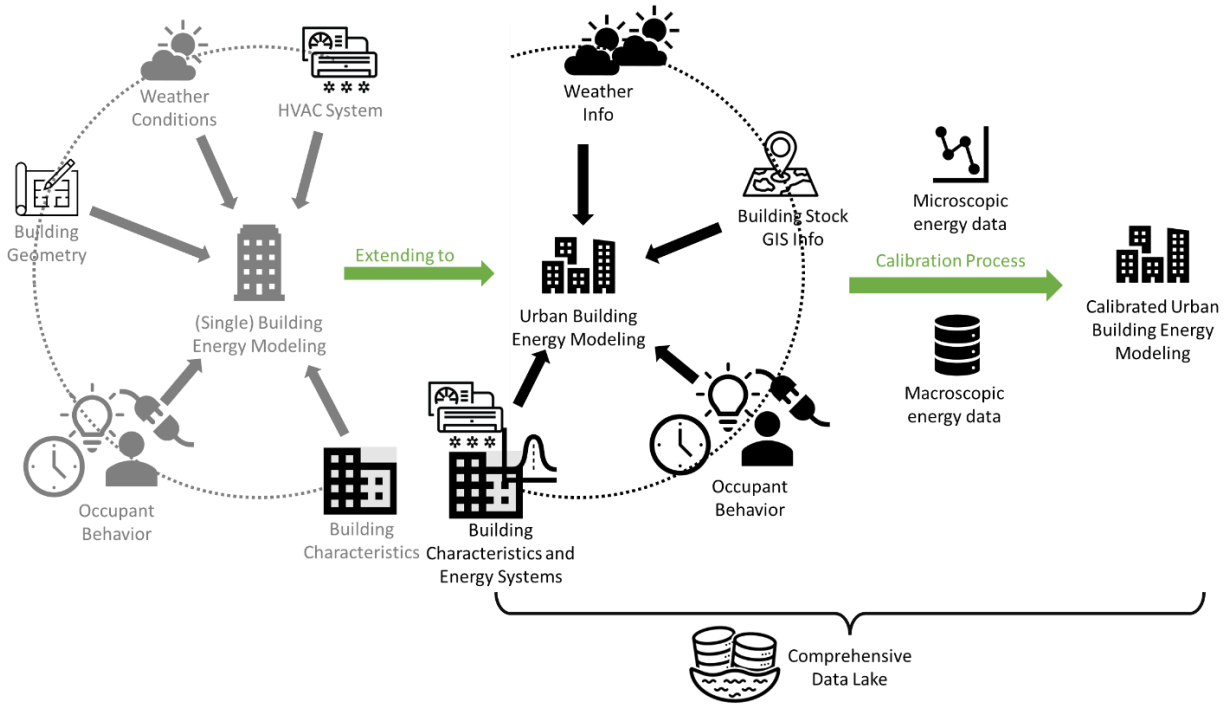

Figure 1. Diagram of inputs of single building energy modeling extended to UBEM

In this section, we aim to highlight the most significant national open data for the readers instead of exhausting all the national open data sources to the reader; besides, we will not review of open data at the local level.

## 2.1. Weather Information

The influence of weather on building energy load underscores its significance as a critical input to physics-based UBEM. Ideally, UBEM necessitates detailed weather data that encapsulates the entirety of the region under study. A higher spatial resolution of weather data enhances the realism of the UBEM.

Open weather-related data primarily fall into two categories. The first category comprises open libraries of Typical Meteorological Year (TMY) and Actual Meteorological Year (AMY) weather data, ready to be used as input by simulation engines like EnergyPlus. A prime example of such open data is found on the EnergyPlus website (https://energyplus.net/weather), which provides weather data for 3,034 locations globally in a format compatible with EnergyPlus. The data includes 1,494 locations in the USA, 80 in Canada, and over 1,450 locations spread across 98 other countries. These weather data sets are systematically organized by World Meteorological Organization region and Country. Additional resources such as Ladybug (https://www.ladybug.tools/epwmap/) and Climate.OneBuilding.Org (https://climate.onebuilding.org/default.html) also offer ready-to-use epw weather files for simulation engines.

The second category of open weather data encompasses datasets from weather services and institutions that are not pre-processed for simulation purposes. These include organizations like NASA (National Aeronautics and Space Administration), National Oceanic and Atmospheric Administration (NOAA), National Centers for Environmental Information (NCEI), and MesoWest. For instance, NASA POWER [18] (https://power.larc.nasa.gov/data-access-viewer/) offers a comprehensive collection of global weather datasets along with a user-friendly API, empowering users to access, analyze, and utilize weather data for various applications. Integrated Surface Database (ISD) [19] collected more than 25,000 ground-based weather stations globally by NOAA and NCEI, which is the primary source for AHSRAE design conditions. Baseline Surface Radiation Network (BSRN) [20] collects meteorological and solar radiation data at 1-to-3-minute resolution with 58 weather stations around the globe. ECMWF ReAnalysis (ERA5) [21] is created by European Centre for Medium-Range Weather Forecast (ECMWF) from multiple satellite and reanalysis datasets. Other open data sets focus on specific weather variables, such as The National Solar Radiation Database (NSRDB) [22] and Photovoltaic Geographical Information System (PVGIS) [23]. Numerous real-time weather data platforms exist as well, such as Weather Underground [24]. The second category of open datasets require additional post-processing by modelers to generate specific weather files for simulation and modeling purposes.

The weather information mentioned above is real-world collected data. Since UBEM often predicts energy based on future weather, synthetic future weather data are essential input to future predictions of energy patterns. The weather data can also be categorized to real-world weather data and weather prediction data. The well-known weather prediction open data include the CMIP (Coupled Model Intercomparison Project) [25], IPCC Assessment Reports [26], CORDEX (Coordinated Regional Climate Downscaling Experiment) [27], etc.

## 2.2. Building Stock GIS Information

In the realm of single-building energy modeling, modelers often rely on detailed floor plans and geometric drawings to construct comprehensive building geometry models, capturing details such as window and door placements and thermal zone divisions. However, replicating this process for UBEM is cost-prohibitive and nearly impossible due to the sheer volume of buildings involved. Instead, the common practice is to capture the external geometry of each building with archetype buildings for UBEM purposes. Building GIS databases provide this crucial external geometry information and, more importantly, include building location coordinates, which allow modelers to select the appropriate weather files for each building's simulation.

The CityGML standard [28], developed by the Open Geospatial Consortium (OGC), provides a conceptual model and exchange format for the representation, storage, and exchange of virtual 3D city models. This aids the integration of urban geodata for various applications related to Smart Cities and Urban Digital Twins (https://www.ogc.org/standards/citygml). CityGML features five consecutive Levels of Detail (LoD), where objects become more detailed with increasing LoD, both in terms of geometry and thematic differentiation. The CityGML standard itself is not open data, but several institutions have adopted the CityGML standard to create open building geometry data. For example, 3D Geoinformation Group from TU Delft offers city data from around the world based on CityGML, which has been widely applied [29]; BuildZero.Org develops the Open City Model project [30] based on CityGML, which seeks to make CityGML data accessible for all U.S. buildings. This entails converting the Microsoft building footprints dataset into the CityGML format, covering LoD1 and LoD2. The project has also integrated additional data on building heights for approximately 10% of the buildings.

In terms of open data released by the industry, Microsoft Maps has released an expansive building footprints dataset for the United States, referred to as the Microsoft footprint (https://github.com/Microsoft/USBuildingFootprints). This dataset comprises 129,591,852 computer-generated building footprints, derived through computer vision algorithms applied to satellite imagery. The data is freely available for download and use. One study has further rasterized the Microsoft footprint data to reduce the computational cost of using it [31]. OpenStreetMap [32] also offers open data and enables users to export data for a selected area. This includes 2D building footprints and a list of tags such as building type, name, and height. Similarly, Google's Open Buildings (https://sites.research.google/open-buildings/) is a large-scale open dataset which contains the outlines of buildings derived from high-resolution satellite imagery in order to support these types of uses. The project is based in Ghana, with an initial focus on the continent of Africa and new updates on South Asia, South-East Asia, Latin America and the Caribbean.

## 2.3. Occupant Behavior

In energy modeling for individual buildings, accurate representation of occupant behaviors is significant since it is influential on various aspects of building operation, i.e., occupancy, HVAC operation, lighting, plug load, etc. Realistically capturing occupant behaviors as inputs of building modeling is important to close the gaps between modeling and measurements [33]. Despite its importance, obtaining occupant behavior data for accurate occupant modeling is challenging due to stochastic and dynamic nature of occupant behaviors, difficulty and cost of behavior monitoring, and privacy concerns in data collection [34] [35] [36]. Currently, most data collection efforts are still limited to a small number of occupants in specific buildings of different regions and climate zones. To overcome these data challenges, researchers have exerted efforts to standardize and organize existing occupant behavior data. Public agencies also conducted national scale surveys periodically to understand human behaviors.

The first category of human behavior data characterizes how occupants respond to environmental signals (e.g., indoor temperature, illuminance level etc.) under different scenarios, represented by the Global Building Occupant Behavior Database [37] [38] and Global Thermal Comfort Database [39] [40] from ASHRAE. The Global Building Occupant Behavior Database comprehensively covers occupancy patterns and human interactions with building systems (e.g., light switching, set-point changes on thermostats, fans on/off, etc.), organized from more than 30 field studies in both commercial and residential buildings from 15 countries. Building on the extended Brick schema, the database records not only occupant behaviors, but also sensor and room metadata information for easier use of users. Meanwhile, related to HVAC control behaviors, the ASHRAE Global Thermal Comfort Database comprises two databases, compiled in 1990s and 2010s respectively, documenting >70k of thermal comfort data from more than hundreds of commercial buildings, residential buildings, and school buildings across the globe. The thermal sensations of occupants, along with occupant background information and environmental factors, are all organized and documented in the database.

Furthermore, capturing occupant behavior patterns and schedules throughout the day is important to reflect the spatiotemporal dynamics of building load profiles in modeling. As a recognized and authoritative dataset maintained by Census Bureau of Labor Statistics, the Time Use Survey (TUS) is widely used to document and quantify the time distribution among individuals for a plethora of activities in weekdays and weekends. Using American Time Use Survey (ATUS) as an example, it documents time allocation and time use patterns of individuals covering a wide range of activities, including work, leisure, household chores, and transportation in different days for a wide spectrum of occupants with diverse socio-demographics, such as age, employment status, and gender, to be representatives of the whole population of US. Relying on ATUS data, researchers can create more realistic occupancy profiles and activity schedules for energy models, thereby enhancing the accuracy and granularity of their energy simulations. As a result, these data can facilitate more reliable energy consumption predictions and optimize decision-making in energy-efficient building design and operation. Another popular Time-of-Use resource is the "Popular Times" feature in Google Maps [41]. As a crucial component of the Google Maps platform, this feature capitalizes on a wealth of location data, providing insights into the popularity of specific places at different times of the day. While assisting individuals in activity planning, crowd avoidance, and travel route optimization, the "Popular Times" function also has substantial potential in enriching various sectors, notably urban planning, transportation management, and building energy modeling.

## 2.4. Building Characteristics (Construction and Energy System) Information

Building characteristics information is the key to the quantification of heating and cooling loads in diverse buildings. Accurate and comprehensive descriptions of these characteristics are pivotal in UBEM. According to Chen et al. [7], essential building characteristic data for UBEM encompass building height, number of stories above and below ground, total and heated floor areas, number of dwellings, construction and refurbishment years, building type, and heating/cooling system type.

We partition the open data related to this context into four categories, collectively referred to as representative sampled building characteristics data. The **first** category comprises representative building data and attributes, often sourced from resources like the Commercial Buildings Energy Consumption Survey (CBECS) [42] and Residential Energy Consumption Survey (RECS). This type of data is collected from survey of various buildings across a certain region. This data sheds light on the physical features, systems, and operational aspects of a wide range of buildings, thereby serving as a groundwork for creating representative or archetype building models and for examining energy consumption patterns on a larger scale.

The **second** category, measured energy performance data, provides an invaluable insight into the actual energy performance and efficiency of individual buildings. The Building Performance Database (BPD) is recognized in the United States as a leading, public compendium of recorded energy performance metrics for numerous residential and commercial structures. The BPD offers in-depth insights into a building's characteristics and energy consumption, encompassing details from structural and system attributes to operational and maintenance practices. This database is instrumental for UBEM as it aids in the formulation of emblematic building energy frameworks. These models are pivotal for energy specialists, city planners, and decision-makers, enabling them to assess potential energy conservation measures and devise strategies for curtailing greenhouse gas emissions at a metropolitan scale. Moreover, the BPD acts as a reference point for gauging building efficiency, helping in pinpointing energy optimization avenues and directing both investment and legislative choices towards bolstering urban energy resilience [43]. In addition to the BPD, the U.S. Department of Energy has initiated the Standard Energy Efficiency Data (SEED) Platform™. SEED stands as an open-source, secure framework for the oversight of extensive building performance metrics from diverse sources. The platform assists municipal and state entities in overseeing building energy data and pinpointing avenues for enhanced energy efficiency within their domains. SEED ensures users can seamlessly amalgamate data from varied origins, sanitize and validate this data, and disseminate the refined information to collaborators. The platform streamlines tasks such as formatting, aligning, sanitizing, and error detection within datasets. It also fosters collaborative data handling, enabling multiple stakeholders to edit a single dataset while maintaining a clear audit trail. Moreover, SEED's application programming interface (API) provides the capability to directly integrate chosen datasets with other software utilities or publicly accessible dashboards [44] [45].

The amount of building energy data available in digital form is increasing rapidly via programs such as city energy disclosure ordinances and audit mandates, the digitization of previously paper-based building transactions, as well as the Internet of Things (IoT). However, the use and utility of this data is not growing as rapidly. One of the core issues is a lack of standardization in terminology and vernacular for quantities as basic as floor area. The Building Energy Data Exchange Specification (BEDES) is a dictionary of terms, definitions, and field formats that was created to help address this problem. BEDES is not a software tool, database or even a schema. It is a dictionary upon which interoperable schema, databases and software tools can be built [46].

**Lastly,** building documentation and records. Building permits records holds a wealth of data including project years, updates to building systems, and changes in building types. Building Permits Survey (BPS) [47] provided national, state, and local statistics on new privately-owned residential construction. The United States Code, Title 13, authorizes this survey, provides for voluntary responses, and provides an exception to confidentiality for public records. Data are available monthly, year-to-date, and annually at the national, state, selected metropolitan area, county and place levels. The second sub-category is building certificate database, which offers information on individual buildings' energy performance and efficiency. The database contains energy performance certificates that detail key factors about a building's energy use, insulation levels, heating/cooling systems, and renewable energy installations. The integration of such open data into UBEM can significantly enhance its accuracy and realism, enabling stakeholders to portray building populations accurately, make informed decisions, design energy-saving interventions, and advocate for sustainable urban growth. Only regional building certificate database is found [48]; no national building certificate database is found in the review.

## 2.5. Macroscopic energy data

The calibration process in UBEM is an essential step to ensure that the models are reliable and accurately reflect real-world energy usage and consumption patterns. To calibrate the model, real-world energy usage data is collected. Once the real-world data has been collected, it's compared to the predictions made by the baseline model. If there are discrepancies, the model's inputs are adjusted to better match the observed data. This can involve adjusting parameters related to equipment efficiency, thermal properties, and occupant behavior, among others. After the model has been adjusted, it's validated by comparing its predictions to a different set of real-world data. This helps ensure that the model isn't just fitted to the initial set of calibration data, but is also able to accurately predict energy use under different conditions. The calibration process is typically iterative. This means that the above two steps are repeated until the model's predictions are within an acceptable range of the observed data. As a result, the macroscopic energy data representing urban-scale total energy consumption of buildings is essential to the calibration processes of UBEM.

Macroscopic energy data representing urban-scale total energy consumption of buildings broadly falls into two categories: data from energy research institution and data from utility company, both of which are instrumental in the calibration process. Data from energy research institution, the first category, includes vital resources such as Energy Information Administration (EIA) and the Buildings Energy Data Book. These sources offer a wealth of insights into large-scale energy consumption patterns, enabling the calibration of models based on representative building data and attributes. Specifically, EIA data provides a holistic view of energy consumption, production, and pricing across various sectors, which facilitates accurate energy demand modeling. The Buildings Energy Data Book, on the other hand, delivers detailed statistics on building energy consumption, helping calibrate models to mirror real-world conditions.

The second category, data from utility company, concentrates on calibrating models to reflect detailed load profiles focusing on the service territory of utility companies. This data, obtained from utility companies, yields granular insights into the energy usage patterns of individual buildings or building clusters. The use of utility company energy data allows researchers to refine their models to accurately depict temporal variations in energy demand and load shapes, resulting in more precise and dependable energy simulations. However, utility data is not always open data, which limits city and county level energy studies.

## 2.6. Microscopic energy data

Microscopic energy data plays a crucial role in the detailed calibration of UBEMs. The term "microscopic energy data" refers to data that is detailed and specific to individual buildings or units, often captured in high-frequency intervals. This type of data provides a granular understanding of energy consumption patterns and enabling precise calibration processes. Microscopic energy data such as the Building Data

Genome Project [49], Reference Energy Disaggregation Data Set (REDD) [50], UK Domestic Appliance-Level Electricity (UK-DALE) [51], and the REFIT dataset [52] impart valuable insights into individual building energy performance. Some datasets also furnish in-depth information about building characteristics, equipment in use, and occupancy besides energy. They provide a framework for comprehensive calibration of models to accurately emulate real-world building dynamics. Beside those well-known microscopic energy data mentioned above, the use of Advanced Metering Infrastructure (AMI) data that are collected from individual buildings can also enhance the calibration process by providing high-resolution energy consumption data at a granular level. AMI data, which includes real-time or interval-based energy measurements, enables researchers to calibrate models with a focus on load profiles and temporal variations in energy demand. By leveraging microscopic energy data from various sources, researchers can achieve a detailed calibration of UBEMs, leading to accurate energy simulations and the identification of energy-saving opportunities at the individual building level.

Furthermore, the value of microscopic energy data is not just confined to the calibration of individual buildings; it extends its significance to the calibration of archetype building energy models. Archetype building energy models are representative models that embody the typical energy behavior and characteristics of a particular category or class of buildings in an urban setting. These archetypes serve as foundational units in the broader scope of UBEM. They bridge the gap between the granular, individual building level and the holistic, city-scale energy modeling. By incorporating microscopic energy data into these archetype models, researchers and urban planners can ensure that these foundational units are calibrated with a high degree of accuracy. Such precision is indispensable because even minor discrepancies in archetype model calibration can amplify when scaled up to city or district levels, leading to significant deviations from real-world energy consumption patterns.

## 2.7. Comprehensive Data Lake

The category of Comprehensive Data Lake refers to a collection of diverse datasets, often stored in raw, unprocessed formats, that can be utilized for UBEM. These data lakes aggregate and offer a vast array of information, useful for model calibration and energy analysis.

One example of such a data repository is the Building Data Genome Project [49]. It offers a broad selection of open datasets encompassing building energy data, weather data, and occupancy data. This variety allows researchers access to a comprehensive range of resources, instrumental for model development and calibration. Data.gov [53] is another notable data lake. It acts as a colossal storehouse of open data, providing numerous datasets pertinent to building energy modeling. Examples of the available data include energy consumption records, weather statistics, and building performance data. In addition to these, other valuable data sources exist, compiled in various online portals. The European Union Open Data Portal [54], the U.S. Census Bureau, and the Duke University Energy Data resources [55], to name a few, present an abundance of datasets related to energy, buildings, climate, and other relevant domains. These resources provide researchers a rich array of data to augment their modeling and analytical endeavors.

The advantage of leveraging these data lakes is the accessibility to additional information for researchers. This extra data can support the UBEM modeling process, model calibration process, validate model outputs, and provide insights into the multifarious factors influencing building energy performance. Thus, comprehensive data lakes contribute significantly to the evolution and improvement of UBEM.

## 2.8 Summary
Among the various facets of open data for UBEM, we examine the present development status of each category and explore the potential avenues for future progression. These insights can offer a roadmap for the evolution of UBEM endeavors.

**Weather Information**. This category stands out as being particularly well-developed, largely attributable to the advancement of satellite-based monitoring systems and machine learning algorithms. These technologies have enhanced the accuracy and forecasting capabilities of weather predictions, making the data more reliable for various applications in urban energy management. **Building Stock GIS Information**. The maturity of Building Stock GIS information owes its development to innovative GIS visualization and mapping technologies. The rise of satellite images and drone surveys and 3D modeling has provided a comprehensive understanding of urban building landscapes, further optimizing energy utilization and policy planning. **Occupant Behavior.** Diverse types of occupant behaviors exist in buildings that affect different aspects of building operation. Despite the challenges in occupant behavior data collection, there is an ongoing trend to organize and open-source existing behavior data from specific studies to deepen the understanding and support modeling of occupant behaviors in general. National survey data represented by time use survey are also valuable data sources to capture occupant behaviors at large and facilitate realistic urban scale building modeling. **Building Characteristics**. A category that requires more concerted attention, building characteristics data often suffer from inconsistencies. The current open data varies in formats, making integration and interpretation tasks daunting. A glaring limitation is the lack of granularity, with a significant dearth of city and county-scale studies. This often results in over-reliance on generalizations and limiting the precision of modeling efforts. The category also requires effort in the development in the cyberinfrastructure, which will open source more building documentation and records data. **Macroscopic Energy Data**. While the broader scale of this data category is commendably developed, there's a conspicuous gap at the city and county scale. This can be attributed to the complexity of recording energy flows at micro levels, compounded by bureaucratic hurdles and infrastructural limitations. **Microscopic Energy Data**. Predominantly in its nascent stage, the main challenge here lies in the proprietary nature of most of the data. Much of the granular energy consumption data is held by private corporations and is utilized for their strategic advantage, making it challenging to access for open research. **Comprehensive Data Lake**. In terms of sheer volume, this category is robust. However, the data often exists in silos, making cross-referencing a challenge. An integrated platform that leverages technologies like semantic web and graph databases can enhance interlinking and promote more comprehensive analyses.

In summary, the well-developed categories include 1) weather information, 2) building stock GIS information, 3) comprehensive data lake, and the underdeveloped categories include 1) occupant behavior, 2) building characteristics, 3) macroscopic energy data (at city and county scale), and 4) microscopic energy data. The assessment of these categories underscores the immense potential and challenges that lie ahead. Bridging the existing gaps can not only refine UBEM's capabilities but can also revolutionize urban scale energy studies.

# 3 Semantic Data Schema for UBEM
## 3.1 Semantic Data Schema
In today's data-rich environment, semantic data schemas have become crucial for effective data handling. These schemas are unique because they focus on the context and relationships in data. This means they don't just organize data points but connect them in meaningful ways. Data schemas simplify data extraction, making it easier to find and use relevant data. Additionally, it aids in better data analysis, helping identify patterns and trends. With the growing challenges in big data and artificial intelligence, semantic data schemas provide a straightforward and scientific method for understanding and utilizing data efficiently. Semantic data schema, or structures that prioritize meaning and context in data organization, emerge as the backbone of efficient and targeted data extraction processes. Within the context of single building, building automation systems employ custom metadata schemas, which define the function and relationships of sensors. However, these schemas differ across buildings and vendors. This inconsistency complicates the use of new applications like energy visualization and anomaly detection, and limits application portability across buildings [56]. To address this, several initiatives such as Project Haystack [57], Brick schemas [58], the Industry Foundation Classes (IFC) [59], the Building Topology Ontology (BOT) [60], and the Smart Appliances REFerence Ontology (SAREF) [61], aiming to create a universal metadata schema. By standardizing schemas, it's easier to apply data-driven techniques, promoting consistent energy management strategies in buildings.

Within the context of urban-scale data, which is the focus of this paper, semantic data schemas also play a pivotal role. For UBEM, where the focus is on urban energy consumption patterns, building characteristics, and their interplay with various external factors, it becomes crucial to filter out the noise from the deluge of open data sources. **From the perspective of efficient data utilization,** many open data repositories carry a wealth of information, much of which may not directly pertain to UBEM. In such scenarios, a generic data extraction method might lead to superfluous data, wasting computational resources and potentially diluting the focus of the research. By emphasizing the extraction of only the most relevant information, the semantic data schema tailored for UBEM can ensure that researchers are equipped with data that is most pertinent to their modeling tasks. This specificity not only streamlines the research process but also enhances the accuracy and precision of the resultant models. Furthermore, as the variety and volume of open data sources continue to expand, maintaining a consistent data interpretation framework becomes paramount. A semantic data schema offers this consistency, ensuring that regardless of the source, data is interpreted, organized, and utilized in a uniform manner, aligned with the objectives of UBEM. **From the perspective of data sharing**, semantic data schema is also important for UBEM. The process of establishing an effective methodology for sharing and utilizing open data in the field of physics-based UBEM necessitates a comprehensive and strategic approach. The community must collaboratively develop processes, principles, and guidelines to address the unique challenges and ethical concerns that arise with open data utilization. The development and implementation of metadata schemas become crucial to assign semantic meanings to the shared data. This provides clarity and facilitates understanding for users accessing the shared information. Reichman et al. [62] underscored the importance of standardizing metadata development. They posited that enhancing the reproducibility of analysis and establishing incentives for data sharing can boost data sharing initiatives. Furthermore, they highlighted the value of well-curated and integrated open data. By addressing these challenges, the community can optimize the utilization of open data in UBEM, fostering more reliable and accurate model development.

Motivated by those observations, semantic data schema for UBEM has been discussed by many researchers due to its importance. Austin et al. [63] suggested a smart city digital twin architecture that encompasses semantic knowledge representation and reasoning. This architecture operates in tandem with machine learning methodologies to facilitate data collection and processing, event identification, and automated decision-making. A real-world example of this approach involved analyzing energy usage in buildings located within the Chicago Metropolitan Area, demonstrating the practical applicability of this framework. Similarly, Bischof et al. [64] are in the process of creating a linked-data model for the semantic annotation of data streams within smart city environments. In line with this effort, the EU FP7 CityPulse project is developing linked-data descriptions for smart city data, further demonstrating the potential for semantically enhanced data structures. Touzani and Granderson [65] studied the domain of building footprint semantic segmentation. They devised a workflow incorporating data pre-processing, deep learning semantic segmentation modeling, and results post-processing, which was then applied to a dataset encompassing remote sensing images from 15 cities and five counties across various regions of the USA. This dataset included an impressive 8,607,677 buildings, demonstrating the large-scale potential of semantic open data in this context. However, many studies including this one only focus on the building geometry semantic segmentation. A comprehensive study that addresses all data types for UBEM is missing.

These endeavors illustrate practical applications for accessing and processing smart city data. They highlight shared attributes that can be adapted for semantic modeling in smart city applications and services. Yet, there remains a significant research gap concerning the application of a systematic semantic data schema tailored for UBEM, especially when relying on open data as the foundational resource. To address this, we introduce "Meta-Urban", a systematic semantic data schema designed specifically for diverse categories of open data in UBEM. The developed semantic data schema Meta-Urban is presented in Table *1*.

Table 1. Developed semantic data schema for UBEM focusing optimized open data utilization

| Data Type | Critical Information and potential component to form meta data schema | Notes |
|---|---|---|
| 1. Weather Information | Name of Dataset/ Website | String value |
|  | Timestamp format | e.g., HH:MM:SS MM/DD/YYYY |
|  | Constant Interval? | True or False |
|  | Time interval | Unit: min, e.g., "1 min", "30 min", "60 min", or "NaN" (if constant interval is True) |
|  | Interpolation Methodology | Linear, etc. |
|  | Location of weather collection | Coordinate in tuple format, e.g. (32.36200946378022, -111.07590759052607) |
|  | Effective coverage | Format such as "GeoJSON" defining the effective area; or geographic information such as country, state, zip code, etc. |
|  | EPW or not | True or False |
|  | AMY/TMY/Others | ["AMY", "TMY", "Others"] |
|  | Weather variable list | e.g., ["Dry-bulb temperature", "Wet-bult temperature", "Direct solar radiation", etc.] |
| 2. Building Stock GIS Information | Name of Dataset/ Website | String value |
|  | Resolution | By building or community level |
|  | 3D or 2D | "3D" or "2D" Building Models |
|  | Data Format | Shapefile/FileGDB, GeoJSON, and CityGML |
|  | Levels of detail (LoD) | 1) LoD1: simple 2-D footprint, 2) LoD2: a box shape, adding slope roofs, 3) LoD3: adding exterior shades and windows and doors, 4) LoD4: and full details of interior layout and zoning |
| 3. Occupant Behavior | Name of Dataset/ Website | String value |
|  | Timestamp format | e.g., HH:MM:SS MM/DD/YYYY |
|  | Constant Interval? | True or False |
|  | Time interval | Unit: min, e.g., "1 min", "30 min", "60 min", or "NaN" (if constant interval is True) |
|  | Building Type | Effective building type list, e.g., ["Office Building", "Single Family House", "Multi-family House" etc.] |
|  | Behavior Type | ["occupancy", "HVAC", "laundry", "cooking", "lighting" etc.] |

| | Control Action | ["Thermostat Adjustment", "HVAC on/off", "Lighting on/off", "Window open/close" etc.] |
|---|---|---|
| | Behavior Relevant Factors | ["Indoor Temperature", "Outdoor Temperature", "Humidity", "Illuminance Level" etc.] |
| | Effective coverage | Format such as "GeoJSON" defining the effective area; or geographic information such as country, state, zip code, etc. |
| 4. Building Characteristics | Name of Dataset/ Website | String value |
| | Effective coverage | Format such as "GeoJSON" defining the effective area; or geographic information such as country, state, zip code, etc. |
| | Sample or Probability | Select from "Sample" and "Probability". The unit of sample is number of buildings; the probability is unitless and its value is less than 1 |
| | Building characteristics distribution list | List all the available building characteristics included in a list, e.g., ["floorspace", "building envelope", "building systems", "principal building activity", "number of story"] |
| | Joint distribution list | Subset of building characteristics distribution list only listing joint distribution groups, e.g., [["floorspace", "number of story"], ["building systems", "principal building activity"]]; if no joint distribution, the value is an empty list [] |
| 5. Macroscopic energy data | Name of Dataset/ Website | String value |
| | Sector | Residential, commercial, all, and unknown |
| | Timeseries | True or False |
| | Timestamp format | e.g., HH:MM:SS MM/DD/YYYY |
| | Constant Interval | True or False |
| | Time interval | e.g., "1 min", "30 min", "60 min", "1 day". "1 month", "1 year", (if constant interval is True), or "NaN" (if constant interval is False) |
| | Interpolation Methodology | Linear, etc. |
| | Effective coverage | Format such as "GeoJSON" defining the effective area; or geographic information such as country, state, zip code, etc. |
| | Energy Data List | e.g., ["Gas", "Electricity", "Primary Energy", "Carbon Emission"] |
| 6. Microscopic energy data | Name of Dataset/ Website | String value |
| | Timestamp format | e.g., HH:MM:SS MM/DD/YYYY |
| | Constant Interval? | True or False |
| | Time interval | Unit: min, e.g., "1 min", "30 min", "60 min", or "NaN" (if constant interval is True) |
| | Interpolation Methodology | Linear, etc. |
| | Geographic information | Geographic information such as country, state, zip code, address, etc. |
| | Existing Metadata Schema? | True or False |

Table *1* segments the data into distinct categories as suggested by the taxonomy developed and introduced in Section 2. Each category is further defined by critical information components and specific metadata schema details for UBEM, aiding in precision and clarity. This table illustrates a comprehensive approach to addressing the diverse data types pertinent to UBEM, showcasing a holistic view of the integral components and their specifications. Through this systematic structure, researchers can better

understand, organize, and utilize data consistently across various UBEM applications. Case studies in Section 3.2 will demonstrate the improved machine-readability and human-interpretability brought by the developed schema.

## 3.2 Schema Alignment: Case Studies

Within UBEM, the implications of effective schema alignment via semantic structures are profound. As models seek to represent real-world energy patterns, behaviors, and influences, any misalignment or discordance in the underlying data can lead to inaccuracies or misconceptions. By leveraging semantic data schema, we ensure that data from various sources—be it about building materials, occupancy patterns, or energy consumption metrics—all align in a manner that reflects machine-readable and human-interpretable urban energy information.

In order to demonstrate the effectiveness of the proposed data schema in Section 3.1, we process 1) Commercial Buildings Energy Consumption Survey (CBECS) data, 2) Energy Information Administration (EIA) monthly energy data, and 3) Residential Energy Consumption Survey (RECS) data as case studies to show how we align their original format to our proposed format, and demonstrate how it will streamline the UBEM process. Specifically, we applied LLM to align the target open data set to the target schema we developed in Section 3.1. LLM can represent the most state-of-the-art AI technology to represent complex linguistic structures, patterns, and relationships within vast amounts of textual data. If LLM works in the case study, it demonstrates that the schema is both human-understandable and machine-readable. Specifically, we applied prompt engineering [66], which is the process of refining and designing input queries to guide the LLM towards desired responses in LLM, to systematically help LLM understand the task. This approach leverages the inherent knowledge and capabilities of the model by providing it with contextually rich prompts, enhancing its ability to generate accurate and task-specific outputs. In terms of LLM, we use ChatGPT-4 code interpreter plug-in under the version of August 3, 2023.

### 3.2.1 Building Characteristics: CBECS

The first case study extracts and aligned data schema from CBECS [42] which represents building characteristics open data. Conducted by EIA, this survey offers detailed insights into factors affecting energy consumption, such as building characteristics, equipment usage, and operational practices. The engineered prompt is shown in Figure *2*.

> Automate everything without asking follow-up questions.
>
> The uploaded File 1 and File 2 are data value and code book. File 2 is optional. Below is the target schema.
>
> "
>
> Name of Dataset/ Website    String value
>
> Effective coverage    Format such as "GeoJSON" defining the effective area; or geographic information such as country, state, zip code, etc.
>
> Sample or Probability    Select from "Sample" and "Probability". The unit of sample is number of buildings; the probability is unitless and its value is less than 1
>
> Building characteristics distribution list   List all the available building characteristics included in a list, e.g., ["floorspace", "building envelope", "building systems", "principal building activity", "number of story"]
>
> Joint distribution list    Subset of building characteristics distribution list only listing joint distribution groups, e.g., [["floorspace", "number of story"], ["building systems", "principal building activity"]]; if no joint distribution, the value is an empty list []
>
> "
>
> The first task is to generate a metadata.csv that represents the aligned schema of the data in file 1 and file 2 in accordance with the schema. Make sure that the rows in the generated csv file are the fields of the schema with one column representing the value.
>
> The second task is to generate processed data under desired format. For each item in the value list (for example "Energy Data List", and "Building characteristics distribution list"), generate a csv file whose first column is timestamp or ID, and second column is value (timeseries or non-timeseries). The first column name is "timestamp" or "ID", and the second column name is the value label.

Figure 2. Engineered prompt framework using Case Study 1 as an example

"File 1" and "File 2" in the designed prompt can be downloaded from the 2018 microdata from CBECS official website. Targe schema is copied and pasted from Table *1*. The resulting aligned schema for CBECS is shown in Figure 3 and the example resulting processed data of CBECS is shown in Figure 4. The meta data is accurate compared with the actual information of CBECS; The processed data is also correct in accordance with the requirement in the prompt.

| Field | Value |
| --- | --- |
| Name of Dataset/ Website | cbecs2018_final_public |
| Effective coverage | Regions: [3 4 1 2], Census divisions: [5 9 7 2 8 3 6 1 4] |
| Sample or Probability | Sample |
| Building characteristics distribution list | ['REGION', 'CENDIV', 'PBA', 'SQFT', 'SQFTC', 'WLCNS', 'RFCNS', 'RFCOOL', 'RFTILT', 'BLDSHP', 'GLSSPC', 'NFLOOR', 'BASEMNT', 'FLCEILHT', 'ATTIC', 'ELEVTR', |
| Joint distribution list | [] |

Figure 3. Resulting Aligned Schema for CBECS

| ID | PBA | | ID | MFBTU |
| --- | --- | --- | --- | --- |
| 1 | 2 | | 1 | 29727152 |
| 2 | 2 | | 2 | 1730655 |
| 3 | 8 | | 3 | 52387 |
| 4 | 5 | | 4 | 3185775 |
| 5 | 5 | | 5 | 4677009 |
| 6 | 14 | | 6 | 3935517 |
| 7 | 14 | | 7 | 2001353 |
| 8 | 5 | | 8 | 3891278 |
| 9 | 25 | | 9 | 2981980 |
| 10 | 14 | | 10 | 24281549 |

Figure 4. Example Resulting Processed Data for CBECS

### 3.2.2 Macroscopic Energy Data: EIA Monthly Energy Data

EIA [67] is a principal agency within the U.S. federal system that collects, analyzes, and disseminates a wealth of information related to energy resources, production, consumption, and distribution. The EIA Monthly Energy Data is a comprehensive dataset, detailing monthly statistics and insights on energy production, consumption, prices, and trends in the United States. We apply the prompt engineering with a similar structure in Section 3.3.1, and the results are also accurate shown in Figure 5 and Figure 6. The aligned schema of EIA Monthly Energy Data perfectly reflects the actual information from EIA, and the correspondent processed data from EIA represent cleaned and correct information.

| Name of Dataset/Website | U.S. Energy Information Administration |
|---|---|
| Sector | Commercial |
| Timeseries | TRUE |
| Timestamp format | YYYY-MM-DD HH:MM:SS |
| Constant Interval | TRUE |
| Time interval | 1 month |
| Interpolation Methodology | Unknown |
| Effective coverage: | U.S. |
| Energy Data List: | Coal Consumed by the Commercial Sector |
| | Natural Gas Consumed by the Commercial Sector (Excluding Supplemental Gaseous Fuels) |
| | Petroleum Consumed by the Commercial Sector (Excluding Biofuels) |
| | Total Fossil Fuels Consumed by the Commercial Sector |
| | Conventional Hydroelectric Power Consumed by the Commercial Sector |
| | Geothermal Energy Consumed by the Commercial Sector |
| | Solar Energy Consumed by the Commercial Sector |
| | Wind Energy Consumed by the Commercial Sector |
| | Biomass Energy Consumed by the Commercial Sector |
| | Total Renewable Energy Consumed by the Commercial Sector |
| | Total Primary Energy Consumed by the Commercial Sector |
| | End-Use Energy Consumed by the Commercial Sector |
| | Total Energy Consumed by the Commercial Sector |

Figure 5. Resulting Aligned Schema for EIA Monthly Energy Data

| Timestamp | Total Primary Energy Consumed by the Commercial Sector |
|---|---|
| 1/1/1973 0:00 | 657.977 |
| 2/1/1973 0:00 | 623.759 |
| 3/1/1973 0:00 | 492.136 |
| 4/1/1973 0:00 | 363.393 |
| 5/1/1973 0:00 | 288.63 |
| 6/1/1973 0:00 | 226.461 |
| 7/1/1973 0:00 | 196.228 |
| 8/1/1973 0:00 | 204.22 |
| 9/1/1973 0:00 | 208.169 |
| 10/1/1973 0:00 | 264.373 |

Figure 6. Example Resulting Processed Data for EIA Monthly Energy Data

### 3.2.3 Occupant Behavior Data: RECS

The third case study is to align and organize occupant behavior related data in RECS with the data schema in Section 3.1. RECS is a national survey collected in a five-years cycle by US EIA to understand the characteristics of residential buildings and occupants at the national level. The surveys documented how occupants use appliances and HVAC systems, including the frequency of using clothes dryer ('DRYRUSE'), dish washer ('DWASHUSE'), temperature setpoint when at home and away ('TEMPHOME', 'TEMPGONE'), etc., to name a few. A similar prompt in Section 3.3.1 was applied. The aligned schema and data are shown in Figure 7 and Figure 8. As the list is long, only a partial list of schema and data is shown here for the demonstration purpose.

| Name of Dataset/Website | RECS 2020 Public Data |
|---|---|
| Timestamp format | NaN |
| Constant Interval? | FALSE |
| Time interval | NaN |
| Behavior Type | ['RCOOKUSE', 'ROVENUSE', 'COOKTOPUSE', 'OVENUSE', 'USECOFFEE', 'DWASHUSE', 'DRYRUSE', 'TVUSE1', 'TVUSE2', 'TVUSE3', 'USEEQUIPAUX', 'USEHUMID', 'USECFAN', 'HOUSEFAN', 'USEDEHUM'......] |
| Control Action | ['WASHTEMP', 'SSLIGHT', 'SSTEMP', 'HEATHOME','TEMPHOME', 'TEMPGONE', 'TEMPNITE', 'COOLCNTL', 'TEMPHOMEAC', 'TEMPGONEAC', 'TEMPNITEAC'......] |
| Behavior Relevant Factors | ['HDD65', 'CDD65'] |
| Study Area | {'Region': ['WEST', 'SOUTH', 'NORTHEAST', 'MIDWEST'], 'Division': ['Mountain South', 'West South Central', 'South Atlantic', 'Middle Atlantic', 'East South Central', 'Pacific', 'West North Central', 'New England', 'East North Central', 'Mountain North']} |

Figure 7. Resulting Aligned Schema for RECS

| DOEID | TEMPHOM | TEMPGON | TEMPHOM | TEMPGONEAC |
|---|---|---|---|---|
| 100001 | 70 | 70 | 71 | 71 |
| 100002 | 70 | 65 | 68 | 68 |
| 100003 | 69 | 68 | 70 | 68 |
| 100004 | 68 | 68 | 72 | 72 |
| 100005 | 68 | 68 | 72 | 72 |
| 100006 | 76 | 76 | 69 | 74 |
| 100007 | 74 | 65 | 68 | 70 |
| 100008 | 70 | 70 | -2 | -2 |
| 100009 | 68 | 60 | 72 | 76 |
| 100010 | 70 | 70 | 74 | 74 |
| 100011 | 72 | 70 | 77 | 77 |
| 100012 | 74 | 74 | 77 | 77 |
| 100013 | 74 | 74 | 74 | 74 |
| 100014 | 70 | 70 | 66 | 66 |

Figure 8. Example Resulting Processed Data for RECS

The case studies show that raw open data can be automatically processed based on the proposed schematic data structure using LLMs. The accurate results generated by LLMs indicate the machine-readability and human-interpretability of the developed semantic data schema.

## 4. Conclusions and Future Work

This paper has presented a detailed exploration of the taxonomy and semantic data schema for open data in Physics-Based UBEM. We have identified the necessity for creating a comprehensive semantic data schema, which is pivotal to streamlining data processes for UBEM and enhancing automation.

We identified the current challenges in handling diverse and inconsistent metadata schemas in open data for UBEM. This underscores the critical need for a common metadata schema, to ensure seamless application porting across buildings and facilitate innovative urban-scale energy modeling and analysis. Our review also highlights the limited research conducted on semantic data structures for open data in UBEM in the context of open data as a primary source. Yet, this exploration revealed promising pathways for open data utilization, improving efficiency and automation in UBEM.

As a result, we developed semantic data schema for each open data category to maintain data consistency and improve model automation for UBEM. We also conduct three case studies to automatically align three well-known open data to the developed schema with LLMs to demonstrate machine-readability and human-interpretability of the schema.

The exciting potential of semantic data structures for open data in UBEM demands further exploration and research. **First**, the developed semantic data schema is biased by authors' limited experience and unexhaustive literature review in UBEM; future work should focus on improving data schema with both sophisticated schema discovery algorithms and expert surveys to make it more generic and widely acceptable within the UBEM community. **Second**, in terms of schema alignment, the case studies only align three open data to the schema; although it achieves high accuracy, the schema alignment needs further investigation across a wider range of diverse open data to further test the both the schema and the alignment technology. Besides, the current case studies have two major limitations: 1) the ability of LLM to handle large documents and complicated schema mapping, e.g., many-to-one, aggregation, pivoting, transpose, and join, is not tested in the case study; 2) how to handle self-consistency issues (e.g., random responses) is not discussed. **Finally**, effective collaboration among researchers, industry experts, and policymakers is pivotal for the successful standardization of semantic data schema of open data for UBEM. The trend is moving towards the development of open-source cyberinfrastructures where individuals can both publish and utilize open data based on a unified schema.